# An Experimental Evaluation of Performance of a Hadoop Cluster on Replica Management

Muralikrishnan Ramane[1], Sharmila Krishnamoorthy[2] and Sasikala Gowtham[3]

[123]Department of Information Technology, University College of Engineering
Villupuram, Tamilnadu, India.
[1]murali.itpro@gmail.com
[2]rksharmila123@gmail.com
[3]gowthamsasi001@gmail.com

## ABSTRACT

*Hadoop is an open source implementation of the MapReduce Framework in the realm of distributed processing. A Hadoop cluster is a unique type of computational cluster designed for storing and analyzing large datasets across cluster of workstations. To handle massive scale data, Hadoop exploits the Hadoop Distributed File System termed as HDFS. The HDFS similar to most distributed file systems share a familiar problem on data sharing and availability among compute nodes, often which leads to decrease in performance. This paper is an experimental evaluation of Hadoop's computing performance which is made by designing a rack aware cluster that utilizes the Hadoop's default block placement policy to improve data availability. Additionally, an adaptive data replication scheme that relies on access count prediction using Langrange's interpolation is adapted to fit the scenario. To prove, experiments were conducted on a rack aware cluster setup which significantly reduced the task completion time, but once the volume of the data being processed increases there is a considerable cutback in computational speeds due to update cost. Further the threshold level for balance between the update cost and replication factor is identified and presented graphically.*

## 1. INTRODUCTION

A distributed system is a pool of autonomous compute nodes [1] connected by swift networks that appear as a single workstation. In reality, solving complex problems involves division of problem into sub tasks and each of which is solved by one or more compute nodes which communicate with each other by message passing. The current inclination towards Big Data analytics has lead to such compute intensive tasks.

Big Data, [2] is termed for a collection of data sets which are large and complex and difficult to process using traditional data processing tools. The need for Big Data management is to ensure high levels of data accessibility for business intelligence and big data analytics. This condition needs applications capable of distributed processing involving terabytes of information saved in a variety of file formats.

Hadoop [3] is a well-known and a successful open source implementation of the MapReduce programming model in the realm of distributed processing. The Hadoop runtime system coupled with HDFS provides parallelism and concurrency to achieve system reliability. The major categories of machine roles in a Hadoop deployment are Client machines, Master nodes and

Slave nodes. The Master nodes supervise storing of data and running parallel computations on all that data using Map Reduce. The NameNode supervises and coordinates the data storage function in HDFS, while the JobTracker supervises and coordinates the parallel processing of data using Map Reduce. Slave Nodes are the vast majority of machines and do all the cloudy work of storing the data and running the computations. Each slave runs a DataNode and a TaskTracker daemon that communicates with and receives instructions from their master nodes. The TaskTracker daemon is a slave to the JobTracker likewise the DataNode daemon to the NameNode.

HDFS [4] file system is designed for storing huge files with streaming data access patterns, running on clusters of commodity hardware. An HDFS cluster has two type of node operating in a master-slave pattern: A NameNode (Master) managing the file system namespace, file System tree and the metadata for all the files and directories in the tree and some number of DataNode (Workers) managing the data blocks of files. The HDFS is so large that replicas of files are constantly created to meet performance and availability requirements.

A replica [5] is usually created so as the new storage location offers better performance and availability for accesses to or from a particular location. In the Hadoop architecture the replica is commonly selected based on storage and network feasibility which makes it fault tolerant so as to recover from failing DataNode. It does replicates files based on a rack aware cluster setup in which by default it replicates each file at three principle locations within the cluster; first copy is stored on local node and the other two copies are stored on remote rack. Additional replicas are stored randomly on any rack which could be configured and overridden using scripts.

The rest of this paper is organized as follows. Section II, discusses related studies on data replication schemes in cluster; Section III describes the proposed system model of a Hadoop cluster and the data locality problem; Section IV evaluates the performance of the system by conducting experiments on varying data replication levels. Finally, Section V concludes and discusses the future scope of this work.

## 2. RELATED STUDIES

The purpose of data replication in HDFS is primarily to improve the availability of data. Replication of a data file serves the purpose of system reliability where if one or more nodes fail in a cluster. Recently, studies were done to improve fault tolerance of data in the presence of failure and few of those are discussed below.

**2.1.**   Abad.C.L, Yi Lu, Campbell.R.H; [6] proposed a data replication and placement algorithm (DARE) that adapts to the fluctuations in workload. It assumes the scheduler is unaware to the data replication policy and was implemented and evaluated using the Hadoop framework. When local data is not available the node retrieves data from a remote nodeand process the assigned task and discards the data after completion. DARE benefits from existing remote data retrievals and selects a subset of the data and creates a replica without consuming additional network and computation resources. Node run the algorithm independently to create replicas that are likely to be heavily accessed. The authors designed a probabilistic dynamic replication algorithm with the following features:

1. Nodes sample assigned tasks and replicates popular files in a distributed manner.

2. Correlated data accesses are distributed over diverse nodes as old replicas deleted and new replicas are created.

Experiments proved 7-times improvement in data locality and 70% improvement in cluster scheduling. Reduces job turnaround time by 16% in dedicated clusters and 19% in virtualized public clouds.

**2.2.** Sangwon Seo, Ingook Jang; [7] proposed optimization schemes such as prefetching and pre-shuffling to solve shared environment problems. Both the above schemes were implemented in a High Performance MapReduce Engine (HPMR). In an Intra block fetching an input split or an intermediate output is prefetched whereas the whole candidate data block is prefetched in the interblock prefetching. The pre-shuffling scheme reduces the amount of intermediate output to shuffle and at the time of pre-shuffling HPMR looks over an input split before the map phase begins and predicts the target reducer where the key-value pairs are segregated. A new task scheduler was designed for pre-shuffling and is used only for the reduce phase. Prefetching schemes improve data locality and Pre-shuffling schemes significantly reduces the shuffling overhead during the reduce phase. The schemes provided following contributions:

1. Performance degradation analysis of Hadoop in a shared MapReduce computation environment.

2. Prefetching and Pre-shuffling schemes to improve MapReduce performance when physical nodes are shared by multiple users.

3. HPMR reduces network overhead and exploits data locality compatible with both dedicated and shared environments.

**2.3.** Khanli.L.M, Isazadeh.A; [8] proposed an algorithm to decrease access latency by predicting the future usage of files. Predictive Hierarchal Fast Spread (PHFS) pre-replicates data in a hierarchal data grid using two phases: collecting data access statistics and applying data mining techniques like clustering and association rule mining all over the system. Files are assigned value α which is between 0 and 1 for representing relationships between files.

Files are arranged according to value of α which is called the PWS (predictive working set). PHFS utilizes the PWS of a file and replicates all members of PWS including the file and all files on the path from source to client. PHFS tries to improve data locality by predicting the user's future demands and pre-replicating them in advance thereby achieving higher availability with optimized usage of resources.

**2.4.** Jungha Lee, JongBeom Lim; [9] proposed a data replication scheme (ADRAP) that is adaptive to overhead, associated with the data locality problem. The algorithm works based on access count prediction to reduce the data transfer time and improves data locality thereby reducing total processing time. The scheme adaptively determines the required replication factor by evaluating data access patterns and recent replication factor for a particular data file. The paper contributes the following:

1. Optimizes replication factor and effectively avoids overhead caused by data replication.

2. Dynamically determines data replication requirements.

3. Minimizes processing time of MapReduce jobs by improving the data locality.

**2.5.** Zaharia.M, Borthakur.D; [10] proposed a delay scheduling method that illustrates the conflict between fairness in scheduling and data locality by designing a fair scheduler for a 600-node Hadoop cluster at Facebook. Delay scheduling schedules jobs according to fairness and waits for a small amount of time letting other jobs to launch tasks. It achieves nearly optimal data locality in a variety of workloads and increases throughput by up to 2x while preserving fairness.

The algorithm is applicable under a wide variety of scheduling policies beyond fair sharing such as the Hadoop Fair Scheduler. HFS has two main goals: Fair sharing and Data locality. To achieve the goal the scheduler reallocates resources between jobs when the number of jobs changes by killing running tasks to make room for the new job and waiting for running tasks to finish.

Delay scheduling performs well in typical Hadoop workloads and is applicable beyond fair sharing. Delay scheduling in HFS is generalized to implement a hierarchical scheduling policy motivated by the needs of Facebook's users. The scheduler divides slots between users based on weighted fair sharing at top-level and allows users to schedule their own jobs using either FIFO or fair sharing.

## 3. PROPOSED WORK

The purpose of this research is to evaluate the performance of the Hadoop cluster and to design a rack aware Hadoop cluster. To achieve the purpose a data replication scheme is adapted to fit the system, implemented using a rack aware Hadoop cluster. In such a cluster tasks are run manually with varying levels of data replication. The setup and the shell scripts required for implementation are presented in detail in the following paragraphs. This research work makes a tiny contribution on;

- Minimizing processing time and data transfer load between racks by improving data locality.

### 3.1. Rack Aware Hadoop Clusters

The Data availability and locality are interrelated domains in the realm of distributed processing which when not handled appropriately leads to performance issues. When a scheduled task is about to run which does not have the data required for it's processing, have to load the data from another node causing poor throughput. This paper deals with similar type situation where cluster of nodes are involved and each node belonging to the same or different rack.

For the purpose of experimental evaluation this paper utilizes a Hadoop cluster setup with one Master node and seven slave nodes each configured manually to define the rack number it belongs. This paper utilizes an improved data placement policy to prevent data loss and improve network performance.

Data blocks are replicated to multiple machines to prevent data loss due to machine failures. In the case of a switch or power failure the NameNode stores information locality of Data Nodes in the network topology and utilizes the information to decide placing data replicas within the cluster. An assumption that two machines in the same rack have more bandwidth and lower latency between each other than two machines in two different racks is considered. It is also assumed cross-rack latency is higher than in-rack latency most of the time. The system is designed for evaluation is shown in figure 1.

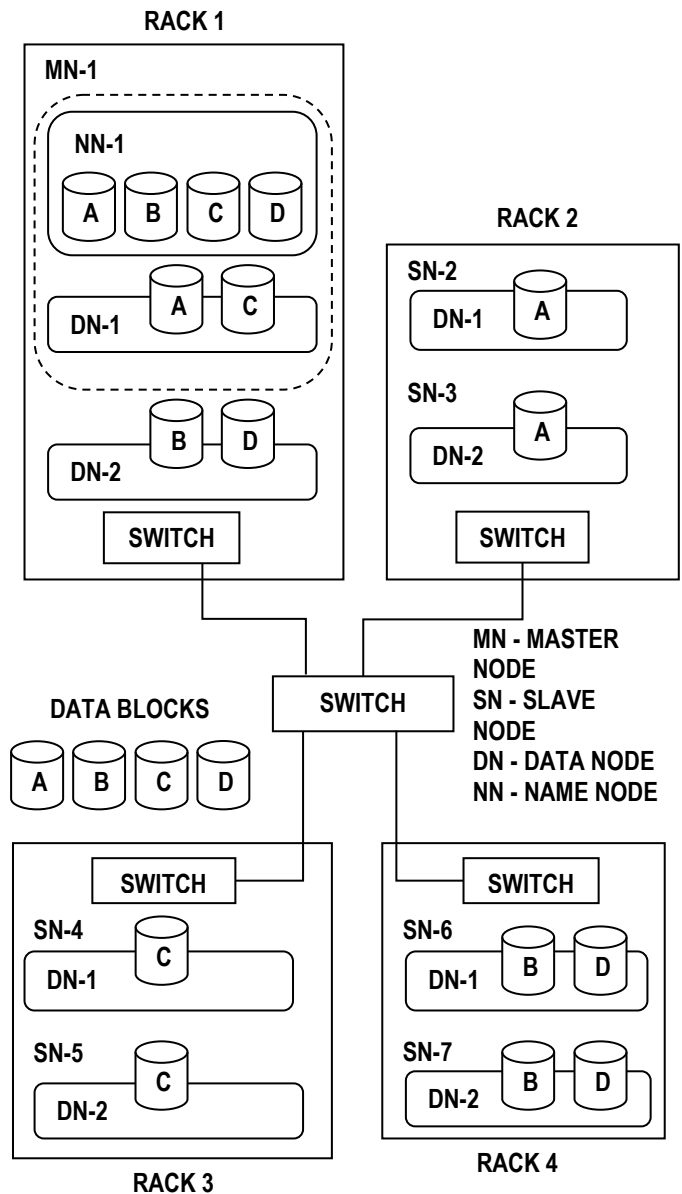

Figure 1.  Rack aware Hadoop Cluster Setup

## 3.2. Access Count Prediction

Maintaining different replication factors per data file and assuming that a higher replication factor for a file with higher access count does not always guarantee better data locality. Data is replicated cautiously for if the replication factor is higher than access count for the particular file then the probability of being processed with node locality is higher than that of the opposite case. To normalize the replication factor, a method that predicts the next access count for a data file is required.

The method initializes the variables then proceeds with calculating the average time interval between data accesses and finally predicts the next access thereby calculating the number of future accesses. The predicted access count is evaluated in comparison with the current replication factor to determine the optimal replication factor. In addition, the number of rack-off locality nodes is effectively reduced by the replica placement policy. To predict access count for individual files this work utilizes Langrange's interpolation using a polynomial expression. The mathematical formula is given below:

$$G(x) = \frac{(x-x_1)(x-x_2)(x-x_3)....(x-x_n)}{(x_0-x_1)(x_0-x_2)(x_0-x_3)....(x_1-x_n)} \, f_0$$

$$+ \frac{(x-x_0)(x-x_2)(x-x_3)....(x-x_n)}{(x_1-x_0)(x_1-x_2)(x_1-x_3)....(x_1-x_n)} \, f_1 ........$$

$$+ \frac{(x-x_0)(x-x_1)(x-x_3)....(x-x_n)}{(x_n-x_1)(x_n-x_2)(x_n-x_3)....(x_n-x_{n-1})} \, f_n \quad ........ (1)$$

In the equation,
Let N be the number of points,
Let $x_i$ be the the i-th point, and
Let $f_i$ be the function of $x_i$.

To calculate the predicted access count,
- substitute x by time t,    where t is time of access and
- y by an access count at t.

### 3.3. Data Placement Policy

In the process of evaluating the data placement policy this work utilized Hadoop clusters that are arranged in racks. In-rack nodes has much more desirable network traffic than off-rack nodes. The cluster administrator uses the configuration variable ***net.topology.script.file.name*** to decide on which rack the nodes belong to and this script is configured so that each node runs the script to determine its rack id. In a default installation nodes are assumed to belong to the same rack with similar rack id.

The data placement policy is designed with a Hadoop cluster in mind which usually is of varying sizes and so varies accordingly. In the case of a small cluster, servers are connected by a single switch with two levels of locality on-machine and off-machine. But for larger installations it must be kept in mind that data replicas exist on multiple machines and spans multiple racks. A rack aware hadoop file system is created with the use of scripts which allows the master node to map the network topology of the cluster. The script is in executable form which allows it to return the rack address of each node.

The script returns the stdout on a list of rack names, one for each input which are provided as arguments such as IP addresses of nodes in the cluster ordered consistently. Mapping scripts specify the key ***topology.script.file.name*** in conf/hadoop-site.xml. The script also provides a command to return rack id's and by default Hadoop will try to send a set of IP addresses as command line arguments. Rack ids are represented as hierarchical path names where every node has a default rack id /default-rack and for large installations rack id's are denoted using the entire topology starting from top level switch to rack names as it is  given here /top-switch-name/rack-name.

### 3.3.1. Configuring Rack Awareness

To configure a Hadoop cluster into a rack-aware system data must be divided into multiple file blocks and store them on different machines among the cluster. In the opposite case, when a Hadoop file system is not rack-aware then there is a possibility that Hadoop will place all the copies of the data blocks in same rack. Failing to configure a Hadoop into a rack-aware system may result in loss of data, but rack failure is not recurrent and this can be avoided by utilizing Hadoop configuration files. Hadoop is configured using the topology property where the actual setup is provided as an input file to the script for rack identification. The following script performs rack identification based on IP addresses given a hierarchical IP addressing scheme enforced by the network administrator.

Configuring rack awareness in Hadoop involves two steps:

- Configure the "topology.script.file.name" in core-site.xml

```
<property>
<name>topology.node.switch.mapping.impl</name>
<value>
org.apache.hadoop.net.ScriptBasedMapping
</value>
</property>
<property>
<name>topology.script.file.name</name>
<value>core/rack-awareness.sh</value>
 </property>
```

**Rack-awareness.sh**

```
HADOOP_CONF=/usr/local/hadoop/conf
while [ $# -gt 0 ] ; do
nodeArg=$1
exec< ${HADOOP_CONF}/topology.data
result=""
while read line ; do
ar=( $line )
if [ "${ar[0]}" = "$nodeArg" ] ; then
result="${ar[1]}"
fi
done
shift
if [ -z "$result" ] ; then
echo -n "/default/rack "
else
echo -n "$result "
fi
done
```

**Topology.data**

```
(master)Machine1.pc        /dc1/rack1
(slave)Machine2.pc         /dc1/rack1
(slave)Machine3.pc         /dc2/rack2
(slave)Machine4.pc         /dc2/rack2
(slave)Machine5.pc         /dc3/rack3
```

| (slave)Machine6.pc | /dc3/rack3 |
| (slave)Machine7.pc | /dc4/rack4 |
| (slave)Machine8.pc | /dc4/rack4 |

## 4. EXPERIMENTAL EVALUATION

Experiments were conducted on the rack-aware Hadoop cluster to evaluate its performance in terms of data availability. It involves two scenarios: one involving too many data files and other with no data files but complex computations. The former one is WordCount application and the latter is Pi value calculation, also both the above experiments were conducted in Hadoop 2.2.0.

Machine Configuration: 8 Nodes (Similar)
Processor: Intel Core i5 3.3 GHZ
RAM: 8 GB
HDD: 400 GB

Nodes within a rack are connected by one Ethernet Switch and one Fast Ethernet switch is used between racks. The size of data block is set to 64Mb with an increasing replication factor starting from 1. Specifically, 8 jobs are run ranging from 1 to 8 replication levels which are not greater than number of nodes available in the cluster. The result of the map phase only is experimented i.e. the completion time and data locality of the map phase is averaged over 8 runs. As noticed, in terms of throughput, the tasks with node locality is better than tasks with rack-off locality.

### 4.1. Results and Graph

Performance evaluations show that as replication levels increase the task completion time gets significantly reduced for computation involving no data files. But for computations that involve data files the completion time reduces and then again shoots up due to update cost. Both experiments were conducted on replication levels ranging from one to eight which is not higher than number of nodes in the cluster.

#### 4.1.1. Experiment for Pi Value

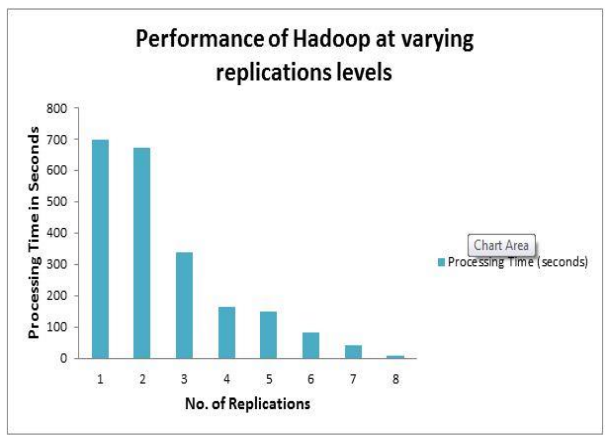

Figure 2. Replication Levels for PI Value

Figure 2 shows that the data replication scheme used in PI value calculation reduces the task completion time. By comparing, with increasing replication factors there is some increase in the performance and when the replication level is increased by 3, its completion time is 337 s, and further reduces considerably to 8.12 s at replication level 8. This shows that as replication factor increases multiple map phases are introduced and thus the computation speeds up.

### 4.1.2. Experiment for WordCount

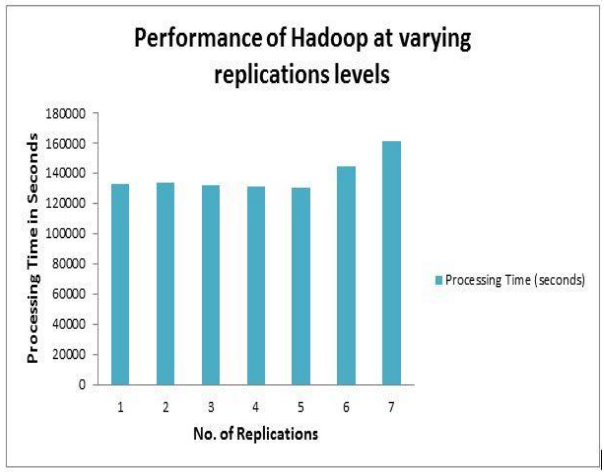

Figure 3. Replication Levels for Word Count

Figure 3 shows the task completion time for WordCount application where the completion time reduces linearly with replication factor increase but once it reaches the threshold level the performance starts to deteriorate. This shows that the computations involving data files do not linearly improve in performance as replication increases. By default, the replication level in HDFS is set to 3 which will reduce the performance speed and thus the completion time is 132220 seconds. On increased replication levels the computation speed boosts up but once it reaches the threshold the time comes down from 1300430 s to 1608600 s.

Performance evaluations show that as replication levels increase the task completion time gets significantly reduced for computation involving no data files. But for computations that involve data files the completion time reduces and then again shoots up due to update cost. Both experiments were conducted on replication levels ranging from one to eight which is not higher than number of nodes in the cluster.

## 5. CONCLUSIONS

Data Replication in Hadoop framework to investigate the data locality problem was experimented and proved that replication improves performance and also decreases it as the threshold limit is crossed. Further the replication factor was supported by an access count prediction algorithm for data files using Lagrange's interpolation which optimizes the replication factor per data file. Performance evaluation showed that our data replication scheme reduces the task completion time. The task completion time for Pi value calculation started with 680 s and came down to 8.12 s and similarly the WordCount application does start with completion time of 1384300 s and came down to 1300430 s. But once the threshold limit is reached the completion time shoots up to 1608600 s.

## ACKNOWLEDGEMENTS

The authors would like to thank all those who have shared their valuable inputs, especially Ganeshpandi and Lakshmi in the Department of Information Technology, University College of Engineering Villupuram, Tamilnadu, India for their insights, suggestions and time throughout the course of this work.

**Authors**

Muralikrishnan Ramane has completed post graduate degree in the field of Computer Science Engineering, specialized in Distributed Computing Systems. He is currently working as a Lecturer at University College of Engineering Villupuram, Villupuram, India. To his credit he has three International Journal and one International Conference publications. His research interests include Data Management in Distributed Environments such as Grid, P2P and Cloud.

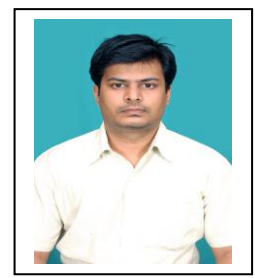